\documentclass[journal=jctcce,manuscript=article]{achemso}
\usepackage{chemformula} 
\usepackage[T1]{fontenc} 
\usepackage{multirow}

\author{Micha\l~Lesiuk}
 \email{lesiuk@tiger.chem.uw.edu.pl}
\author{Bogumi\l~Jeziorski}
 \affiliation{ 
 Faculty of Chemistry, University of Warsaw\\
 Pasteura 1, 02-093 Warsaw, Poland
 }

\title[Riemann zeta-function extrapolation]{Complete basis set extrapolation of electronic correlation energies 
using the Riemann zeta function}


\begin{document}

\begin{tocentry}
\includegraphics[scale=0.62]{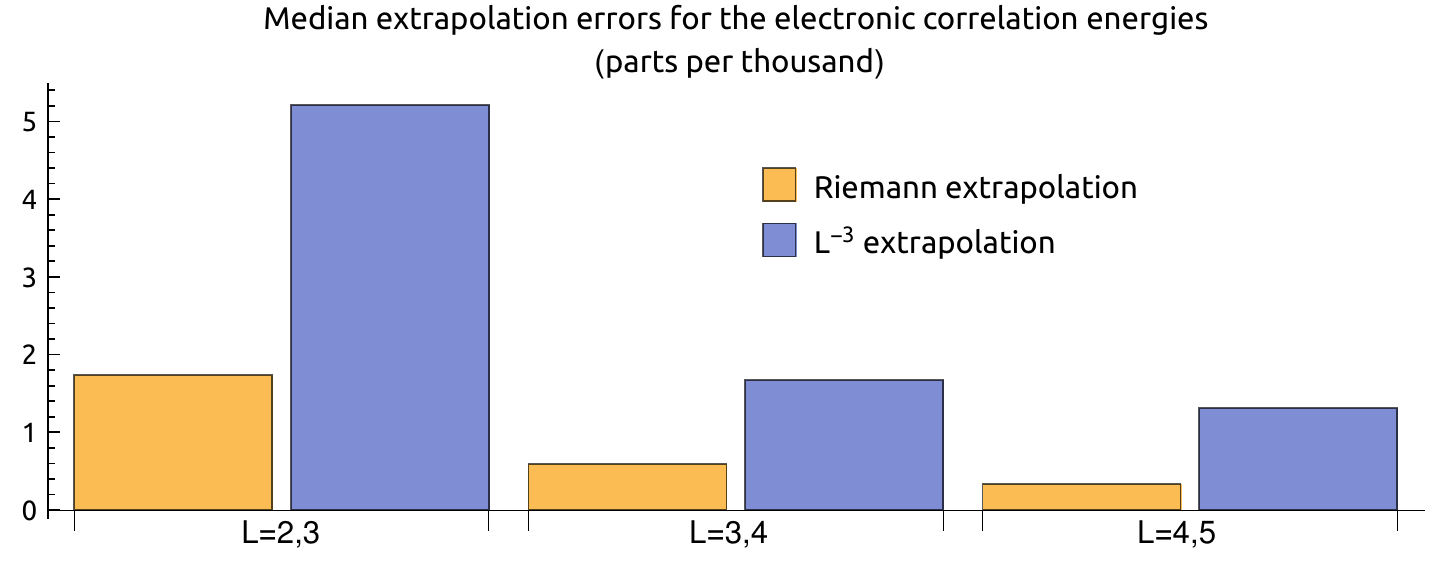}
\end{tocentry}

\begin{abstract}
In this communication we demonstrate the effectiveness of the method of complete basis set (CBS) extrapolation of 
correlation energies based on the application of the  Riemann zeta function. Instead of fitting the results obtained 
with a systematic sequence of one-electron bases with a certain functional form, an analytic re-summation of the 
missing contributions coming from higher angular momenta, $l$ is performed. 
The assumption that these contributions vanish asymptotically as an inverse power of $l$ leads to an expression for the 
CBS limit given in terms of the zeta function. This result is turned into an extrapolation method that is very easy to 
use and requires no ``empirical'' parameters to be optimized. The performance of the  method is assessed by 
comparing the results with very accurate reference data obtained with explicitly correlated theories and with results 
obtained with standard extrapolation schemes. On average, the errors of the zeta-function extrapolation are several 
times smaller compared with the conventional schemes employing the same sequence of bases. A recipe for the estimation 
of the residual extrapolation error is also proposed.
\end{abstract}

\section{Introduction}
\label{sec:intro}

Extrapolation to the complete basis set (CBS) limit is an effective theoretical tool that allows to remove a 
significant fraction of the finite basis set error at essentially zero computational cost. This is illustrated well by 
the papers of Feller and collaborators \cite{feller11,feller13} who have shown that even the worst-performing 
extrapolation schemes are able to reduce the error of raw \emph{ab initio} values by approximately a factor of two.

Various extrapolation protocols have been proposed in the literature (see Refs. 
\cite{klopper01,varandas07,feller11,varandas18} for a comprehensive review). One of the first formulas used for 
this purpose is a simple two-parameter exponential \cite{feller92,feyer96}, later generalized to a three-parameter 
mixed 
exponential/Gaussian expression \cite{peterson94,woon94a}. Both of them are based on purely empirical arguments and it 
has been shown that they typically underestimate the CBS limit \cite{feller06}. 

A majority of the extrapolation schemes used today assume that the results converge towards the CBS limit with the error 
decaying as an inverse 
power of the the largest angular momentum $L$ included in the basis set. This approach, grounded in numerical 
observations of Schwartz \cite{schwartz62a,schwartz62b} for the energy of the helium atom, was subsequently justified 
theoretically by Hill \cite{hill85} and other authors \cite{carroll79,kutz92,kutz08}.

One of the most popular extrapolation schemes employed for the correlation energy is due to Helgaker et al. 
\cite{helgaker97,halkier98} and relies directly on the $L^{-3}$ error formula. However, numerous generalizations and 
extensions of this scheme were proposed in the literature, e.g., the $L^{-\alpha}$ or $(L+1/2)^{-\alpha}$ formulae
with an adjustable $\alpha$ \cite{martin96,martin97,jeziorska03}, expressions that additionally include higher inverse 
powers of $L$ \cite{varandas00,varandas07}, or shifted formulae where some effective parameter is used instead of $L$
\cite{bakowies07,varandas14,pansini15}. A different approach was proposed by Schwenke \cite{schwenke05} who calculated 
the CBS limit by scaling the difference between the results obtained in two largest basis sets. The scaling 
coefficient is determined ``empirically'' for a given level of theory and basis set family. Note that 
extrapolations are not limited to the correlation energy and are also frequently applied to other quantities such as 
polarizabilities \cite{jun08} or structural parameters \cite{puzza09}.

In this work we consider a CBS extrapolation method that is fundamentally different from the 
approaches described in the previous paragraphs. The performance of the proposed method is assessed by 
comparing the results with accurate reference data and with results 
obtained with standard extrapolation schemes. Finally, we put forward a simple recipe to estimate the residual 
extrapolation error.

\section{Theory}
\label{sec:theory}

Let us denote the energy calculated with a basis set including functions up to the angular momentum $L$ by $E_L$. The 
basic idea of the  method advocated  by us relies on the assumption that energy increments, $\delta_l = E_l -E_{l-1}$, 
possess the 
following asymptotic form
\begin{eqnarray}
\label{dxasym}
 \delta_l \rightarrow a\cdot l^{-4} \;\;\; \mbox{as} \;\;\; l \rightarrow \infty,
\end{eqnarray}
where $a$ is a system-dependent numerical parameter. We adopt a convention $E_{-1}=0$, so that $\delta_0 = E_0$.  The 
formula (\ref{dxasym}) is asymptotically valid for spin-singlet electron pairs. A more general (and presumably more 
accurate) asymptotic expression takes also the triplet electron pairs into account and reads
\begin{equation}
\label{dxasym2}
 \delta_l \rightarrow a\cdot l^{-4}+b\cdot l^{-6} \;\;\; \mbox{as} \;\;\; l \rightarrow \infty.
\end{equation}
Here we concentrate mostly on Eq. (\ref{dxasym}), but at the end of this section the main results are generalized to 
take Eq. (\ref{dxasym2}) into account. Note that for systems where all electron pairs are triplet-coupled, such as 
the quintet state of the helium dimer \cite{przybytek05}, one should set $a=0$ in Eq. (\ref{dxasym2}).

The exact energy, $E_\infty$, can obviously be obtained by summing all energy increments and so one can  split this 
summation 
into two parts
\begin{eqnarray}
 E_\infty = \sum_{l=0}^{L} \delta_l + \sum_{l=L+1}^{\infty} \delta_l = 
 E_L + \sum_{l=L+1}^{\infty} \delta_l,
\end{eqnarray}
where $L$ is the maximal angular momentum that is feasible in practice. It is  assumed that $L$ is large enough to make 
Eq. 
(\ref{dxasym}) a good approximation to $\delta_l$ for $l>L$. Under these conditions the CBS limit can be rewritten 
as
\begin{eqnarray}
\label{ezeta0}
 E_\infty = E_L + a\sum_{l=L+1}^{\infty} l^{-4},
\end{eqnarray}
or
\begin{eqnarray}
\label{ezeta}
 E_\infty = E_L + a\Big[\zeta(4)-\sum_{l=1}^{L} l^{-4}\Big],
\end{eqnarray}
where $\zeta(s)=\sum_{n=1}^\infty n^{-s}$ is the Riemann zeta function, so that $\zeta(4)=\pi^4/90$. The 
parameter $a$ required in Eq. (\ref{ezeta}) 
can be fixed assuming that results from two consecutive basis sets, say, $E_L$ and $E_{L-1}$, are available. The 
optimal $a$ is then found straightforwardly as
\begin{eqnarray}
\label{azeta}
 a = L^4\,\big(E_L-E_{L-1}\big).
\end{eqnarray}
Together with Eq. (\ref{ezeta}) this constitutes the basic two-point extrapolation scheme. 

In the above analysis we have assumed that the values of $E_l$ obtained within a given basis set are exact. This is not 
strictly true in practice since the number of $s$,~$p$,~$d$,~$\ldots$, functions is always finite. Further in this work 
we do not assume that each component $E_l$ is radially saturated with high accuracy like, e.g., in the work of  
Moncrieff and Wilson \cite{moncrief98}, and expect the validity of the extrapolation for the conventional correlation 
consistent bases when the number of radial functions decreases with $l$. The error in $E_l$ usually converges rather 
quickly (at least for atoms) with the number of functions $n$ in the shell especially when the ``radial'' functions 
satisfy the nuclear cusp condition \cite{klopper86,kutz08b}. The problem of radial incompleteness is also alleviated by 
using uncontracted and/or augmented basis sets.

When the results from three consecutive basis sets, $E_L$, $E_{L-1}$, and $E_{L-2}$, are available, the three-point 
extrapolation scheme based on Eq. (\ref{dxasym2}) can be applied. In this case the expression for the 
CBS energy limit reads
\begin{eqnarray}
\label{ezeta2}
 E_\infty = E_L + a\Big[\zeta(4)-\sum_{l=1}^{L} l^{-4}\Big] + b\Big[\zeta(6)-\sum_{l=1}^{L} 
l^{-6}\Big],
\end{eqnarray}
where $\zeta(6)=\pi^6/945$, and the coefficients $a$, $b$ are
\begin{eqnarray}\label{azeta2}
 a &=& \frac{L^6\,\big(E_L-E_{L-1}\big)-(L-1)^6\,\big(E_{L-1}-E_{L-2}\big)}{2L-1},\\\label{bzeta2}
 b &= &L^6\,\big(E_L-E_{L-1}\big) - aL^2.
\end{eqnarray}

In principle, for a sufficiently large $L$ the three-point extrapolation should outperform the two-point method. 
The former not only accounts for the convergence of the correlation energy for the triplet electron pairs, but 
the $l^{-6}$ term also serves as a subdominant term in the asymptotic expansion for the singlet pairs. On the other 
hand, the main disadvantage of the three-point formula is the fact that the results from the smallest $L-2$ basis set 
may not be reliable enough. Additionally, the three-point formula may be more susceptible to over-parametrization, 
numerical noise, and 
small irregularities in the basis set sequence.

The method considered  here 
bears some similarities to the convergence acceleration technique employed by Drake and Yan 
\cite{drake95} in  calculations of three-electron atomic integrals over exponential functions.  Its particular variant,
Eqs. (\ref{ezeta}) and (\ref{azeta}), 
has also been applied by Moncrieff and Wilson \cite{moncrief98} to extrapolate the second-order (M\o ller-Plesset) 
correlation energies  for 
14-electron diatomics computed using specially designed universal Gaussian basis sets.  
 
In the next section the results obtained with help of Riemann extrapolations are compared with the following 
conventional formulas \cite{helgaker97,halkier98}
\begin{eqnarray}
\label{helga2}
 E_L & =& E_\infty + A\cdot L^{-3},\\
\label{helga3}
 E_L &= &E_\infty + A\cdot L^{-3} + B\cdot L^{-5}.
\end{eqnarray}
The parameters $E_\infty$, $A$, $B$ are adjusted to match the results obtained with a sequence of 
two or three basis sets. The  extrapolation formula $ E_L = E_\infty + A(L+1/2)^{-4}  + A(L+1/2)^{-6} $ proposed by 
Martin \cite{martin96} has also 
been tested, but we  have  found it inferior in comparison to Eqs.~(\ref{helga2}) and (\ref{helga3}).

It is also interesting to compare the two-point zeta-function extrapolation with the conventional formula 
(\ref{helga2}) under the assumption that Eq. (\ref{dxasym}) strictly holds for each $l\geq L$. In such case the 
zeta-function formula gives the exact result given by Eqs. (\ref{ezeta0}) or (\ref{ezeta}). Employing the asymptotic 
expansion \cite{paris05} of the infinite sum in Eq. (\ref{ezeta0}), the limit $E_\infty$ can be rewritten as $E_\infty 
= 
E_L + \frac{1}{3}a\cdot L^{-3}-\frac{1}{2}a\cdot L^{-4}+\mathcal{O}(L^{-5})$. The extrapolation formula 
(\ref{helga2}) also can be rewritten in an analogous form, $E_\infty' = E_L -a\cdot 
L^{-4}\,[1-L^3\,(L-1)^{-3}]^{-1}$ which for large $L$ can be expanded as $E_\infty' = E_L + \frac{1}{3}a\cdot 
L^{-3}-\frac{2}{3}a\cdot L^{-4}+\mathcal{O}(L^{-5})$. By combining the obtained expressions for $E_\infty$ and 
$E_\infty'$ one can show that the error of the conventional extrapolation, $\epsilon_L=E_\infty-E_\infty'$, behaves 
asymptotically as $\epsilon_L = \frac{1}{6}a\cdot L^{-4} + \mathcal{O}(L^{-5})$. Thus, the 
extrapolation (\ref{helga2}) contains the $L^{-4}$ error coming from inexact summation of the 
$l^{-4}$ term. This error is absent in the zeta-function method which can be viewed as a 
mathematical rationale behind the improved effectiveness of the zeta-function extrapolations. 

\section{Numerical results}
\label{sec:numbers}

\subsection{Correlation energies}

To investigate the performance of the   extrapolation scheme defined by Eqs. (\ref{ezeta})-(\ref{bzeta2}),    we 
performed finite 
basis set calculations for several systems where benchmark-quality results at various levels of theory are 
known. Literature sources of the reference values (which are all accurate to $1\,\mu$H or better) and 
basis sets used in this work are summarized in Table I. Uncontracted Gaussian-type basis sets are used 
throughout the present work. For neon atom we employ Slater-type orbitals (STOs) basis sets optimized for the purposes 
of this work according to the scheme detailed in Refs. \cite{lesiuk15,lesiuk17}. All basis sets can be obtained 
from the authors upon request. The internuclear distance in H$_2$ molecule is set to $1.4\,$a.u. and the geometry of 
H$_3^+$ is the equilateral triangle with the side length of $1.65\,$a.u. All orbitals were active in the correlated 
calculations except for the neon atom where $1s$ core orbital was frozen. Calculations reported in this subsection 
were performed with help of the \textsc{Gamess} program package\cite{gamess1}. 

The results of the two-point and three-point zeta-function extrapolations for the helium atom, beryllium atom (STOs 
basis set), carbon atom, hydrogen molecule (basis set of Mielke \emph{et al.}\cite{mielke99}), and trihydrogen cation 
(H$_3^+$) at the FCI level of theory are given in Table~\ref{tab:fci}. Analogous results at the MP2 and CCSD levels of 
theory for the beryllium atom and lithium hydride molecule (LiH) are summarized in Table~\ref{tab:mp2}. In 
Table~\ref{tab:triples} we show values of 
the perturbative triples correction obtained for the neon atom
\begin{eqnarray}
 E_{\mbox{\scriptsize (T)}} &= E_{\mbox{\scriptsize CCSD(T)}} - E_{\mbox{\scriptsize CCSD}},
\end{eqnarray}
where $E_{\mbox{\scriptsize CCSD}}$ and $E_{\mbox{\scriptsize CCSD(T)}}$ are the correlation energies obtained from the 
CCSD \cite{purvis82} and CCSD(T) methods \cite{ragha89}, respectively. For comparison, the corresponding results 
obtained at the MP2 level of theory are also reported in Table~\ref{tab:triples}.

The results given in Tables~\ref{tab:fci} and \ref{tab:mp2} indicate a good performance of both the two-point and 
the three-point zeta-function extrapolations. On average, the errors are reduced by a factor of 3--4 compared with the 
conventional schemes employing the same number of basis functions. The gains are usually larger for smaller basis sets; 
for 
example, the quality of the results obtained even with the most crude $L=2,3$ zeta-function extrapolation compares 
favourably with raw results from basis sets as large as $L=5,6$. In some cases the three-point zeta-function method 
does 
not lead to any improvement over its two-point counterpart. This may be due to increased susceptibility of the former 
to 
small irregularities in the basis set sequence or simply is an indication that the convergence with respect to the 
angular momentum has already been reached and other factors are limiting the accuracy at this point (such as the 
``radial'' convergence).

Similar conclusions apply also to the results for the neon atom presented in Table~\ref{tab:triples}. For the 
$E_{\mbox{\scriptsize (T)}}$ component of the correlation energy the two-point $L=6,7$ zeta-function extrapolation 
outperforms any other scheme giving an error of only about $0.002\,$mH (which is very close to the estimated accuracy 
of the benchmark value).

In many applications, especially in those that aim at high accuracy of the results, it is important to provide a 
reliable estimate for the error of the extrapolated value. The simplest solution is to take 
the difference between the extrapolated value and the result in the largest basis set available. This conservative 
approach has been used in the literature numerous times \cite{patkowski15,przybytek17}, but in the case of 
the zeta-function method it leads to gross overestimations. Let us consider a more general 
approach where the residual extrapolation error, $\sigma_L=|E_{\mbox{\scriptsize exact}}-E_{\infty}|$, is estimated as
\begin{eqnarray}
\label{sigl}
 \sigma_L=C_L|E_L-E_{\infty}|,
\end{eqnarray}
where $C_L$ is a numerical parameter. By using the results from 
Tables~\ref{tab:fci}-\ref{tab:triples} we can find ``empirical'' values of $C_L$ that can be used to estimate 
the error of future applications. To this end, we demand that for each extrapolated value from 
Tables~\ref{tab:fci}-\ref{tab:triples} the 
error calculated from Eq. (\ref{sigl}) is not smaller than the true error. In the case of the two-point zeta-function 
extrapolation one finds
\begin{eqnarray*}
 &C_3 = 0.113, \;\;\;\;\;\; C_4 = 0.135, \;\;\;\;\;\; C_5 = 0.136, \;\;\;\;\;\; C_6 = 0.094, \;\;\;\;\;\; C_7 = 0.094,
\end{eqnarray*}
Note that the fact that $C_3<C_4$ does not imply that the $L=2,3$ extrapolation gives, on average, 
smaller errors than $L=3,4$ extrapolation. The latter extrapolation would still be significantly more reliable simply 
because $|E_4-E_{\infty}|$ is much smaller than $|E_3-E_{\infty}|$, cf. Eq. (\ref{sigl}). The error bars obtained with 
help of the constants $C_L$ can be additionally tightened if the value of $\sigma_L$ is interpreted statistically as 
the standard deviation, i.e., one demands that in about 68\% of cases the true error is smaller than $\sigma_L$, in 
about 95\% of cases the true error is smaller than $2\sigma_L$, and so forth (assuming that the error distribution is 
normal). However, a further justification of this approach requires a larger statistical sample than available here 
and shall be considered in 
future works. One may note that another popular method of estimating the 
extrapolation error, namely taking the difference between the last two extrapolated results \cite{lesiuk15}, was found 
by us to underestimate the error in several cases and we do not recommend its use in combination with the zeta-function 
method.

\subsection{Other quantities}

To check the applicability of the zeta-function method to extrapolation of properties other than the energy we 
selected static dipole polarizabilities of the helium atom and of the 
hydrogen molecule, and the exchange splitting in the hydrogen molecule.
The latter quantity is defined as a difference between the energies of the ground ($^1\Sigma_g^+$) and the first 
excited 
state ($^3\Sigma_u^+$) of this system. Here we consider a scaled quantity defined as\cite{pachucki10}
\begin{eqnarray}
\label{split}
 \Delta E = e^{2R} R^{-5/2} (E_u - E_g),
\end{eqnarray}
where $E_g$ and $E_u$ are the energies of the ground and excited states, respectively, and $R$ is the internuclear 
distance ($R=8.0$ was adopted in the calculations). The static dipole polarizabilities were calculated with help 
of the Dalton program package \cite{dalton,christian98}. Doubly-augmented Gaussian-type basis sets of Woon and 
Dunning \cite{woon94b} were used in polarizability calculations for the hydrogen molecule. 

The data provided in Tables~\ref{tab:hepol} and \ref{tab:exch} reveal that extrapolation of properties is a 
considerably more difficult task than of the correlation energies. Nonetheless, zeta-function extrapolations 
still give considerably better results than the conventional extrapolations with the same number of points. The only 
exception is the two-point extrapolation for the helium atom, but even in this case the differences are marginal.
Overall, the extrapolated values converge less regularly to the CBS limit compared with the results for the correlation 
energies discussed in the previous section. This suggests that the lack of radial saturation may be responsible for the 
remaining basis set incompleteness error, despite doubly-augmented basis sets were used in the calculations.

\section{Conclusions}
\label{sec:conclusion}

We have studied the performance  of the complete basis set extrapolation that is based on analytic re-summation of 
the missing energy increments using the Riemann zeta function. The performance of the proposed method has been assessed 
by comparing with accurate 
reference data obtained with explicitly correlated theories and with results obtained with standard extrapolation 
schemes. For extrapolation of the correlation energies we recommend the simplest two-point zeta-function formula due to 
its reliability, ease of use, and regular convergence of the results. This scheme outperforms the standard 
extrapolation methods in most cases studied here and, on average, allows to reduce the residual extrapolation errors 
several times. We have also proposed a reliable method of estimating the remaining extrapolation error.

In the future this work can be extended in several directions. For example, separate treatment of singlet and triplet 
electron pairs can be performed, and an analogous separation can be made  for the core and valence electron pairs. It is 
also 
possible to generalize the zeta-function method to extrapolate relativistic and quantum electrodynamics corrections 
that are known to converge pathologically slowly \cite{kutz08,balcerzak17} with respect to the basis set 
size.

\begin{acknowledgement}
The authors thank prof. K. Patkowski for fruitful discussions and for reading and commenting on the manuscript. This 
work was supported by the 
National Science Centre, Poland, within the project 2017/27/B/ST4/02739.
\end{acknowledgement}

\begin{table}[b]
\caption{\label{tab:ref} Literature sources of benchmark values and of basis sets employed in this 
work. The abbreviation HF stands for Hartree-Fock.}
\begin{tabular}{cccccc}
\hline\hline
 & benchmark & basis set \\
\hline\\[-2.0ex]
\multirow{4}{*}{He}      & Nakashima \emph{et al.} \cite{nakashima08} (FCI limit)     
                         &  \\
                         & Lehtola \cite{lehtola18} (HF limit) & Cencek \emph{et al.} \cite{cencek12} \\
                         & Pachucki and Sapirstein \cite{pachucki00} & (d$X$Z) \\
                         & (static polarizability) & \\[0.5ex]
\hline\\[-2.0ex]
\multirow{4}{*}{Be}      & Przybytek \emph{et al.} \cite{przybytek18} 
                         & Prascher \emph{et al.} \cite{prascher11} \\
                         & (MP2 and CCSD limits)             & (aug-cc-pwCV$n$Z) \\
                         & Pachucki and Komasa \cite{pachucki04} 
                         & Lesiuk \emph{et al.} \cite{lesiuk19} \\
                         & (FCI limit)             & (Slater-type basis set) \\[0.5ex]
\hline\\[-2.0ex]
\multirow{2}{*}{C}       & Strasburger \cite{strasburger19a,strasburger19b} (FCI limit) & Kendall \emph{at 
al.}\cite{kendall92} \\
                         & Bunge \cite{bunge93} (HF limit) & (aug-cc-pCVXZ) \\
\hline\\[-2.0ex]
\multirow{2}{*}{H$_2$}   & Pachucki \cite{pachucki10} (FCI limit) & Mielke \emph{et al.}\cite{mielke99} \\
                         & Mitin \cite{mitin00} (HF limit)        & (aug-mcc-pV$X$Z) \\
                         & Rychlewski \cite{rychlewski80}         & Woon and Dunning \cite{woon94b} \\
                         & (static polarizability)                & (d-aug-cc-pV$X$Z) \\[0.5ex]
\hline\\[-2.0ex]
\multirow{2}{*}{Ne}      & K\"{o}hn \cite{kohn10} (CCSD(T) limit) & this work \\
                         & Flores \cite{flores08} (MP2 limit) & (Slater-type basis set) \\
\hline\\[-2.0ex]
\multirow{2}{*}{LiH}     & Bukowski \emph{et al.} \cite{bukowski99} 
                         & Prascher \emph{et al.} \cite{prascher11} \\
                         & (MP2 and CCSD limits)             & (aug-cc-pwCV$n$Z) \\[0.5ex]
\hline\\[-2.0ex]
\multirow{2}{*}{H$_3^+$} & Pavanello \emph{et al.} \cite{pava09} (FCI limit) & Mielke \emph{et al.} \cite{mielke99} \\
                         & Jensen \emph{et al.} \cite{jensen05} (HF limit)   & (aug-mcc-pV$X$Z) \\[0.5ex]
\hline\hline
\end{tabular}
\end{table} 
\newpage

\begin{table}
\caption{\label{tab:fci} Extrapolation errors (in $\mu$H) in the FCI correlation energy for the helium atom, beryllium 
atom, carbon atom, hydrogen molecule (H$_2$), and trihydrogen cation (H$_3^+$). The values of $-E_L$ (in mH) are given 
in the second column.
The reference values of the correlation energy are $E_\infty(\mbox{He})=-42.044\,381\,$mH, 
$E_\infty(\mbox{Be})=-94.332\,459\,$mH, $E_\infty(\mbox{C})=-156.287\,\,$mH, 
$E_\infty(\mbox{H$_2$})=-40.846\,348\,$mH, and $E_\infty(\mbox{H$_3^+$})=-43.463\,500\,$mH.}
\begin{tabular}{ccccccc}
\hline\hline
$L$ & $-E_L$ & Eq. (\ref{ezeta}) & Eq. (\ref{helga2}) & Eq. (\ref{ezeta2}) & Eq. (\ref{helga3}) \\
\hline\\[-2.0ex]
\multicolumn{6}{c}{He} \\
\hline\\[-2.0ex]
 2  & 40.018\,397 & ---             & ---            & ---             & --- \\
 3  & 41.173\,663 & 171.0           & 384.3          & ---             & --- \\
 4  & 41.597\,808 & \phantom{0}58.8 & 137.1          & 21.4            & 82.9 \\
 5  & 41.785\,680 & \phantom{0}27.2 & \phantom{0}61.6& 11.3            & 32.1 \\
 6  & 41.881\,296 & \phantom{0}14.4 & \phantom{0}31.7& \phantom{0}5.6  & 14.7 \\
 7  & 41.934\,921 & \phantom{00}8.6 & \phantom{0}18.3& \phantom{0}3.5  & \phantom{0}8.0 \\
\hline\\[-2.0ex]
\multicolumn{6}{c}{Be} \\
\hline\\[-2.0ex]
 2  & 85.976\,344 & ---                & ---   & ---             & --- \\
 3  & 91.479\,502 & $-$383.7           & 632.4 & ---             & --- \\
 4  & 92.994\,102 & \phantom{$+$0}50.2 & 329.6 & 194.6           & 263.3 \\
 5  & 93.608\,566 & \phantom{$+$0}63.4 & 175.8 & \phantom{0}70.1 & 115.7 \\
 6  & 93.902\,091 & \phantom{$+$0}70.5 & 123.7 & \phantom{0}75.5 & \phantom{0}93.9 \\
\hline\\[-2.0ex]
\multicolumn{6}{c}{C} \\
\hline\\[-2.0ex]
 2  & 132.539\,255 & ---               & ---    & ---  & --- \\
 3  & 145.933\,543 & 2240.9            & 4713.9 & ---  & --- \\
 4  & 151.028\,728 & \phantom{0}600.1  & 1540.3 & 53.8 & 844.9 \\
\hline\\[-2.0ex]
\multicolumn{6}{c}{H$_2$} \\[0.4ex]
\hline\\[-2.0ex]
 2  & 39.834\,097 & ---             & ---            & ---                          & --- \\
 3  & 40.449\,439 & 24.2           & 137.8           & ---                          & --- \\
 4  & 40.652\,767 & \phantom{0}7.7 & \phantom{0}45.2 & \phantom{$-$}2.2             & 24.9 \\
 5  & 40.737\,378 & \phantom{0}4.7 & \phantom{0}20.2 & \phantom{$-$}3.2  & 10.4 \\
 6  & 40.779\,706 & \phantom{0}0.8 & \phantom{00}8.5 & $-$1.8  & \phantom{0}1.8 \\
\hline\\[-2.0ex]
\multicolumn{6}{c}{H$_3^+$} \\[0.5ex]
\hline\\[-2.0ex]
 2  & 42.370\,983 & ---    & ---              & ---       & --- \\
 3  & 43.051\,539 & $-$0.2 & 125.4            & ---       & --- \\
 4  & 43.271\,252 & $-$8.6 & \phantom{0}31.9  & $-$11.4   & 11.4 \\
 5  & 43.354\,056 & $+$7.4 & \phantom{00}22.6 & $+$15.5   & 18.9 \\
 \hline\hline
\end{tabular}
\end{table}
\newpage

\begin{table}
\caption{\label{tab:mp2} Extrapolation errors (in $\mu$H) in the MP2 and CCSD correlation energies for the beryllium 
atom and for the lithium hydride (LiH) molecule. All abbreviations are the same as in Table \ref{tab:fci}. The 
reference values of the MP2 correlation energy are $E_\infty(\mbox{Be})=-76.358\,249\,$mH and 
$E_\infty(\mbox{LiH})=-72.889\,5\,$mH, and of the CCSD correlation energy are $E_\infty(\mbox{Be})=-93.664\,5\,$mH 
and $E_\infty(\mbox{LiH})=-82.990\,1\,$mH.}
\begin{tabular}{ccccccc}
\hline\hline
$L$ & $-E_L$ & Eq. (\ref{ezeta}) & Eq. (\ref{helga2}) & Eq. (\ref{ezeta2}) & Eq. (\ref{helga3}) \\
\hline\\[-2.0ex]
\multicolumn{6}{c}{$E_{\mbox{\scriptsize MP2}}$(Be)} \\
\hline\\[-1.5ex]
 2  & \phantom{0}63.574\,069 & ---              & ---              & ---                & --- \\
 3  & \phantom{0}70.880\,240 & 1052.8           & 2401.7           & ---                & --- \\
 4  & \phantom{0}73.672\,818 & \phantom{0}132.3 & \phantom{0}647.6 & $-$174.2           & 263.3 \\
 5  & \phantom{0}74.849\,721 & \phantom{00}58.5 & \phantom{0}273.7 & \phantom{0$+$}21.2 & 127.8 \\
\hline\\[-2.0ex]
\multicolumn{6}{c}{$E_{\mbox{\scriptsize MP2}}$(LiH)} \\
\hline\\[-1.5ex]
 2  & \phantom{0}55.999\,799 & ---                & ---              & ---                & --- \\
 3  & \phantom{0}66.006\,642 & $+$821.9           & 2669.5           & ---                & --- \\
 4  & \phantom{0}69.669\,934 & $-$129.6           & \phantom{0}546.4 & $-$446.4           & \phantom{0}81.2 \\
 5  & \phantom{0}71.128\,900 & \phantom{0}$-$36.9 & \phantom{0}229.9 & \phantom{0}$-$10.0 & 106.4 \\
\hline\\[-2.0ex]
\multicolumn{6}{c}{$E_{\mbox{\scriptsize CCSD}}$(Be)} \\
\hline\\[-1.5ex]
 2  & \phantom{0}84.563\,558 & ---                & ---             & ---                & --- \\
 3  & \phantom{0}90.325\,338 & $-$149.1           & 914.7           & ---                & --- \\
 4  & \phantom{0}92.237\,729 & $-$320.1           & \phantom{0}32.8 & $-$377.0           & $-$160.5 \\
 5  & \phantom{0}92.895\,696 & \phantom{0}$-$40.3 & \phantom{0}80.0 & \phantom{$+$}101.3 & \phantom{$+$0}98.4 \\
\hline\\[-2.0ex]
\multicolumn{6}{c}{$E_{\mbox{\scriptsize CCSD}}$(LiH)} \\
\hline\\[-1.5ex]
 2  & \phantom{0}68.961\,499 & ---                & ---                & ---                & --- \\
 3  & \phantom{0}78.630\,548 & $-$1496.9          & \phantom{$+$}288.3 & ---                & --- \\
 4  & \phantom{0}81.216\,330 & $-$590.4           & $-$113.3           & $-$288.6           & $-$201.2 \\
 5  & \phantom{0}81.902\,675 & \phantom{$+$}241.7 & \phantom{$+$}367.2 & \phantom{$+$}662.9 & \phantom{$+$}554.8 \\
 \hline\hline
\end{tabular}
\end{table}
\newpage

\begin{table}
\caption{\label{tab:triples} Extrapolation errors of the $E_{\mbox{\scriptsize (T)}}$ and $E_{\mbox{\scriptsize MP2}}$ 
components of the correlation energy for the neon atom. All values are given in mH and the abbreviations are the same 
as in Table \ref{tab:fci}. The reference results are $E_{\mbox{\scriptsize (T)}}=-6.501\,$mH and 
$E_{\mbox{\scriptsize MP2}}=-320.223\,$mH.}
\begin{tabular}{ccccccc}
\hline\hline
$L$ & $E_L$ & Eq. (\ref{ezeta}) & Eq. (\ref{helga2}) & Eq. (\ref{ezeta2}) & Eq. (\ref{helga3}) \\
\hline\\[-2.0ex]
\multicolumn{6}{c}{$E_{\mbox{\scriptsize (T)}}$} \\[0.2ex]
\hline\\[-1.5ex]
 2  & $-$4.277 & ---              & ---                & --- & --- \\
 3  & $-$5.788 & $-$0.202 & 0.077 & ---                & --- \\
 4  & $-$6.185 & $-$0.048 & 0.026 & \phantom{$-$}0.003 & \phantom{$-$}0.014 \\
 5  & $-$6.336 & $-$0.021 & 0.007 & $-$0.007           & $-$0.000 \\
 6  & $-$6.401 & $-$0.002 & 0.010 & \phantom{$-$}0.010 & \phantom{$-$}0.011 \\
 7  & $-$6.435 & $+$0.002 & 0.009 & \phantom{$-$}0.006 & \phantom{$-$}0.008 \\
 \hline\\[-2.0ex] 
 \multicolumn{6}{c}{$E_{\mbox{\scriptsize MP2}}$} \\
\hline\\[-1.5ex]
 2  & 246.555 & ---                & ---                 & --- & --- \\
 3  & 287.239 & \phantom{$+$}8.341 & 15.853              & ---                & --- \\
 4  & 303.478 & \phantom{$+$}1.899 & \phantom{0}4.895    & $-$0.247           & 2.494 \\
 5  & 310.424 & \phantom{$+$}1.240 & \phantom{0}2.510 & \phantom{$+$}0.907 & 1.580 \\
 6  & 313.898 & \phantom{$+$}0.922 & \phantom{0}1.551 & \phantom{$+$}0.703 & 1.002 \\
 7  & 315.887 & \phantom{$+$}0.596 & \phantom{0}0.953 & \phantom{$+$}0.310 & 0.498 \\
 \hline\hline
\end{tabular}
\end{table}
\newpage

\begin{table}
\caption{\label{tab:hepol} Extrapolation errors (in $\mu$H) in the static dipole polarizability of the helium atom 
and the hydrogen molecule ($R=1.4$) calculated at the FCI level of theory. The values of the polarizability calculated 
in the basis set $L$ are given in the second column. The reference values are 
$\alpha(\mbox{He})=1.383\,192\,174\,455(1)\,$a.u. and 
$\alpha_{||}(\mbox{H}_2)=6.387\,318\,8\,$a.u.}
\begin{tabular}{ccccccc}
\hline\hline
$L$ & $\alpha_L$ & Eq. (\ref{ezeta}) & Eq. (\ref{helga2}) & Eq. (\ref{ezeta2}) & Eq. (\ref{helga3}) \\
\hline\\[-2.0ex] 
\multicolumn{6}{c}{$\alpha(\mbox{He})$} \\
\hline\\[-1.5ex]
 2  & 1.385\,972 & ---                 & ---                 & ---                  & --- \\
 3  & 1.384\,154 & $-$138.7            & $+$196.9            & ---                  & --- \\
 4  & 1.383\,522 & $-$247.8            & $-$131.1            & $-$284.1             & $-$203.0 \\
 5  & 1.383\,314 & $-$134.7            & \phantom{0}$-$96.6  & \phantom{0}$-$77.5   & \phantom{0}$-$83.1 \\
 6  & 1.383\,244 & \phantom{0}$-$57.7  & \phantom{0}$-$45.0  & \phantom{00}$-$4.8   & \phantom{0}$-$15.4 \\
 7  & 1.383\,216 & \phantom{0}$-$27.3  & \phantom{0}$-$22.4  & \phantom{00}$-$0.7   & \phantom{00}$-$5.2 \\
\hline\\[-2.0ex]
\multicolumn{6}{c}{$\alpha_{||}(\mbox{H}_2)$} \\
\hline\\[-1.5ex]
 1  & 6.463\,948 & ---                 & ---                    & ---                 & --- \\
 2  & 6.407\,661 & \phantom{$-$}2490.1 & \phantom{$-$}12\,301.6 & ---                 & ---                \\
 3  & 6.394\,838 & \phantom{0}$-$247.6 & \phantom{0$-$}2119.9   & \phantom{$-$}1374.4 & \phantom{$-$}729.4 \\
 4  & 6.390\,487 & \phantom{0}$-$809.4 & \phantom{0000}$-$6.6   & \phantom{0}$-$472.5 & \phantom{$-$}996.4 \\
 5  & 6.389\,139 & \phantom{0$-$}159.3 & \phantom{00$-$}405.8   & \phantom{0$-$}566.8 & \phantom{}$-$649.6 \\
 \hline\hline
\end{tabular}
\end{table}
\newpage

\begin{table}
\caption{\label{tab:exch} Scaled exchange splitting in the hydrogen molecule ($R=8.0$) defined according to Eq. 
(\ref{split}) calculated at the FCI level of theory. All values are given in the atomic units and the abbreviations are 
the same as in Table \ref{tab:fci}. The reference value is $\Delta E=1.736\,967\,949(9)$a.u.}
\begin{tabular}{ccccccc}
\hline\hline
$L$ & $\Delta E$ & Eq. (\ref{ezeta}) & Eq. (\ref{helga2}) & Eq. (\ref{ezeta2}) & Eq. (\ref{helga3}) \\
\hline\\[-2.0ex]
 2  & 1.6825 & ---       & ---       & ---       & --- \\
 3  & 1.6999 & $-$0.0266 & $-$0.0298 & ---       & --- \\
 4  & 1.7142 & $-$0.0096 & $-$0.0123 & $-$0.0040 & $-$0.0084 \\
 5  & 1.7203 & $-$0.0091 & $-$0.0102 & $-$0.0089 & $-$0.0094 \\
 6  & 1.7246 & $-$0.0058 & $-$0.0065 & $-$0.0034 & $-$0.0044 \\
 \hline\hline
\end{tabular}
\end{table}

\bibliography{ref_extra}

\end{document}